\documentclass[conference]{IEEEtran}
\ifCLASSINFOpdf
\else
\fi
\usepackage{graphicx}
\usepackage{url}
\usepackage{float}

\begin{document}
%
\title{Semantic Property Graph for Scalable Knowledge Graph Analytics}

\author{\IEEEauthorblockN{Sumit Purohit}
\IEEEauthorblockA{Pacific Northwest National Laboratory\\
Richland, WA 99352\\
Sumit.Purohit@pnnl.gov}
\and
\IEEEauthorblockN{Nhuy Van}
\IEEEauthorblockA{Pacific Northwest National Laboratory\\
	Richland, WA 99352\\
	Nhuy.Van@pnnl.gov}
\and
\IEEEauthorblockN{George Chin}
\IEEEauthorblockA{Pacific Northwest National Laboratory\\
	Richland, WA 99352\\
	George.Chin@pnnl.gov}
}


%


\maketitle

\begin{abstract}
Graphs are a natural and fundamental representation of describing  the  activities,  relationships,  and  evolution of various complex systems. Many domains such as communication, citation, procurement, biology, social media, and transportation can be modeled as a set of entities and their relationships. Resource Description Framework (RDF) and Labeled Property Graph (LPG) are two of the most used data models to encode information in a graph. Both models are similar in terms of using basic graph elements such as nodes and edges but differ in terms of modeling approach, expressibility, serialization, and target applications. RDF is a flexible data exchange model for expressing information about entities but it tends to a have high memory footprint and inefficient storage, which does not make it a natural choice to perform scalable graph analytics. In contrast, LPG has gained traction as a reliable model in performing scalable graph analytic tasks such as sub-graph matching, network alignment, and real-time knowledge graph query. It provides efficient storage, fast traversal, and flexibility to model various real-world domains. At the same time, the LPGs lack the support of a formal knowledge representation such as an ontology to provide automated knowledge inference. We propose Semantic Property Graph (SPG) as a logical projection of reified RDF into LPG model. SPG continues to use RDF ontology to define type hierarchy of the projected graph and validate it against a given ontolgoy. We present a framework to convert reified RDF graphs into SPG using two different computing environments. We also present cloud-based graph migration capabilities using Amazon Web Services. 
\end{abstract}


%
\IEEEpeerreviewmaketitle

\section{Introduction}
Graphs have been widely used to represent information about entities and their relationships with other entities in any complex system. Many real-world domains that involve activities among a set of entities can be represented as a large graph or a sequence of small graphs \cite{barabasi2000scale,kumar2010structure, cottam2018multi}. A Knowledge Graphs (KG) \cite{paulheim2017knowledge, liuqiao2016knowledge,suchanek2008yago} is a specialized graph constructed to represent the common knowledge about the world as we see it. A KG is a formal and structured representation of facts, relationships, and semantic descriptions of a set of entities. Recent advances in automation on Internet of Things (IoT), social networks, and artificial intelligence-based complex systems have led to a significant increase in the data generation volume. Additionally, it has also highlighted the need for improved interoperability between different components of a system. Domain-specific KGs are also developed to represent entities, relationships, and processes that are characteristic of the domain. They help to combine disparate datasets from different sources, improve data quality by providing metadata to data, and are used in developing domain-specific analytic solutions. Elements of a KG such as facts, relationships, entities, and constraints are defined using an Ontology. An Ontology is a formal specification of concepts. An ontology defines a type hierarchy to specify different relationships in a graph. 

The basic abstraction used to describe a system is considered the \textit{model} used to define the system. Any model defines key elements of the system and relationships between them. Resource Description Framework (RDF) \cite{berners2001semantic, rdfprimer11} and Labeled Property Graph (LPG) \cite{rodriguez2010constructions} are two of the most used data models to encode information in a graph. Both models are similar in terms of using basic graph elements such as nodes and edges but differ in terms of the modeling approach, expressibility, serialization, and target applications. RDF is a flexible data exchange model for expressing information about entities. State-of-the-art Natural Language Processing (NLP) tools use standard ontologies to generate large scale KGs using the RDF data model. Such RDF graphs tend to have a high memory footprint and inefficient storage, which does not make them a natural choice to perform scalable graph analytics. In contrast, LPG has gained traction as a reliable model in performing scalable graph analytic tasks such as sub-graph matching, network alignment, and real-time knowledge graph queries. It provides efficient storage, fast traversal, and flexibility to model various real-world domains. At the same time, the LPGs lack the support of a formal knowledge representation such as an ontology to provide automated knowledge inference. We propose the Semantic Property Graph (SPG) as a logical projection of reified RDF into the Property Graph model. 

The rest of the paper is organized to present our research work. Section \ref{sec:graphdatamodel} presents a brief introduction of RDF and LPG data models and presents key differences. Section \ref{sec:spg} introduces SPG and its key properties. A generation framework is defined in Section \ref{sec:spggen} and Section \ref{sec:spgusecase} presents a Graph analytics use case describing a political conflict scenario. 

\section{Graph Data Models}\label{sec:graphdatamodel}
A \textit{Data Model} is a collection of conceptual tools used to model representations of real-world  entities  and  the relations  among  these  entities \cite{silberschatz1996data}.  Codd \cite{codd1980data} presents that a data model consists of three components: a set of data structure types, a set of operators or inference rules, and a set of integrity rules. Angles et al. \cite{angles2008survey} characterizes a graph data model as that in which data structures for the schema and instances are modeled as graphs or generalizations of them, and the data manipulation is expressed by graph-oriented operations and type constructors. Overall, the graph data model provides some advantages over relational data models. It provides a level of abstraction that allows more natural modeling of information \cite{angles2012comparison}. It also naturally models relationships between data points (or entities) and provides a structure to the information, which is critical to gain insight from the data. The structure also allows intuitive query construction and ad-hoc pattern mining \cite{choudhury2017nous}. 

The RDF \cite{berners2001semantic, rdfprimer11} and LPG \cite{rodriguez2010constructions} are two of the most used graph data models. The RDF is a framework for expressing information about resources \cite{rdfprimer11}. Resources can be anything, including people, location, events, physical objects, and abstract concepts. RDF provides ways to express and interchange machine-readable information without loss of meaning. RDF is used to publish and interlink data on the web using LinkedData concepts \cite{linkeddata}. An RDF statement consists of three elements called triple. The \textit{triple} has a $<subject> <predicate> <object>$ structure. The $\textit{subject}$ and the $\textit{object}$ represent the two resources being related and \textit{predicate} (also called \textit{property}) represents the nature of their directional relationship. We omit a lot of technical details about RDF for brevity and would encourage readers to access RDF 1.1 Primer \cite{rdfprimer11}. RDF is a flexible data exchange model for expressing information about entities. RDF supports complete atomic decomposition of information
because of its \textit{triple} structure. The \textit{$<subject>$} and \textit{$<object>$} are represented as nodes in the graph and the \textit{$<predicate>$} is represented as an edge. All three are uniquely identified by a Uniform Resource Identifier (URI) and are the \textit{atomic} structure of the RDF graph, i.e., they cannot have their internal structure. Every information about them is described using another set of triples. RDF is a World Wide Web Consortium (W3) standard and can take advantage of many other standards and ontologies.  It can also be serialized into different formats such as N-Triple, N3, Turtle, and JSON-LD, etc. SPARQL is the query language used to query RDF data once it is loaded into an RDF compliant graph database. 

The LPG is another graph data model used to describe, store, explore, and graphically depict graph data. The LPG also uses nodes and edges to describe the entities and relationships between them. In contrast to the RDF model, nodes and edges in LPG are not the atomic elements. Instead, they can have an internal structure that represents corresponding \textit{properties} modeled as a key-value pair. This leads to a compact data representation and a more intuitive graph structure. It also provides efficient storage and fast graph traversal, which is essential for many graph analytic applications. The LPG also supports multi-edges of the same type between two nodes. This feature is not part of the RDF specification, without changing the data model and identifying individual instances using different URIs. RDF uses \textit{Reification} to define \textit{Statement} nodes, which are used to describe data about the relationship. This meta graph adds another layer of modeling concepts on top of the input graph. \textit{Reification} gives the required flexibility to describe additional information but this also leads to slower graph traversal and a significant increase in the serialization size of the graph.

\section{Semantic Property Graph} \label{sec:spg}
Network analytics is a multidisciplinary research domain that combines graph gheory, network science, statistics, and machine learning principles to extract insight from a given complex system, modeled as a graph. Many real-world systems can be modeled as a graph and analyzed using different graph algorithms and techniques. Graph traversal and pattern matching are some of the basic graph operations. Given a collection of networks, an important problem is to identify the correspondences across the vertex sets of the networks.  This problem, known as the graph matching problem, has different applications across fields as diverse as computer vision, social network analysis, network de-anonymization and privacy, and biology.  The problem is computationally complex and is known to be NP-hard.  Similarly, network alignment is the task of identifying corresponding nodes in different networks. It has applications across the social, biological, and natural sciences. Many state-of-the-art approaches make use of standard graph properties such as degree distribution, clustering coefficient, and diameter, etc., to model the target system. The graph model (ex: RDF or LPG) used for the underlying input graph does influence the measurement of these graph properties. Similarly, a graph modeled in both RDF and LPG  exhibits different values and distributions for a given property as shown in Section \ref{sec:spgusecase}. Graph visualization and graph summarization are also examples of research tasks that are directly affected by the underlying graph model. This paper focuses on the network analytic task but the framework can also be used for these tasks.

\begin{figure*}
	\centering
	\includegraphics[width=.7\textwidth,height=.35\textheight]{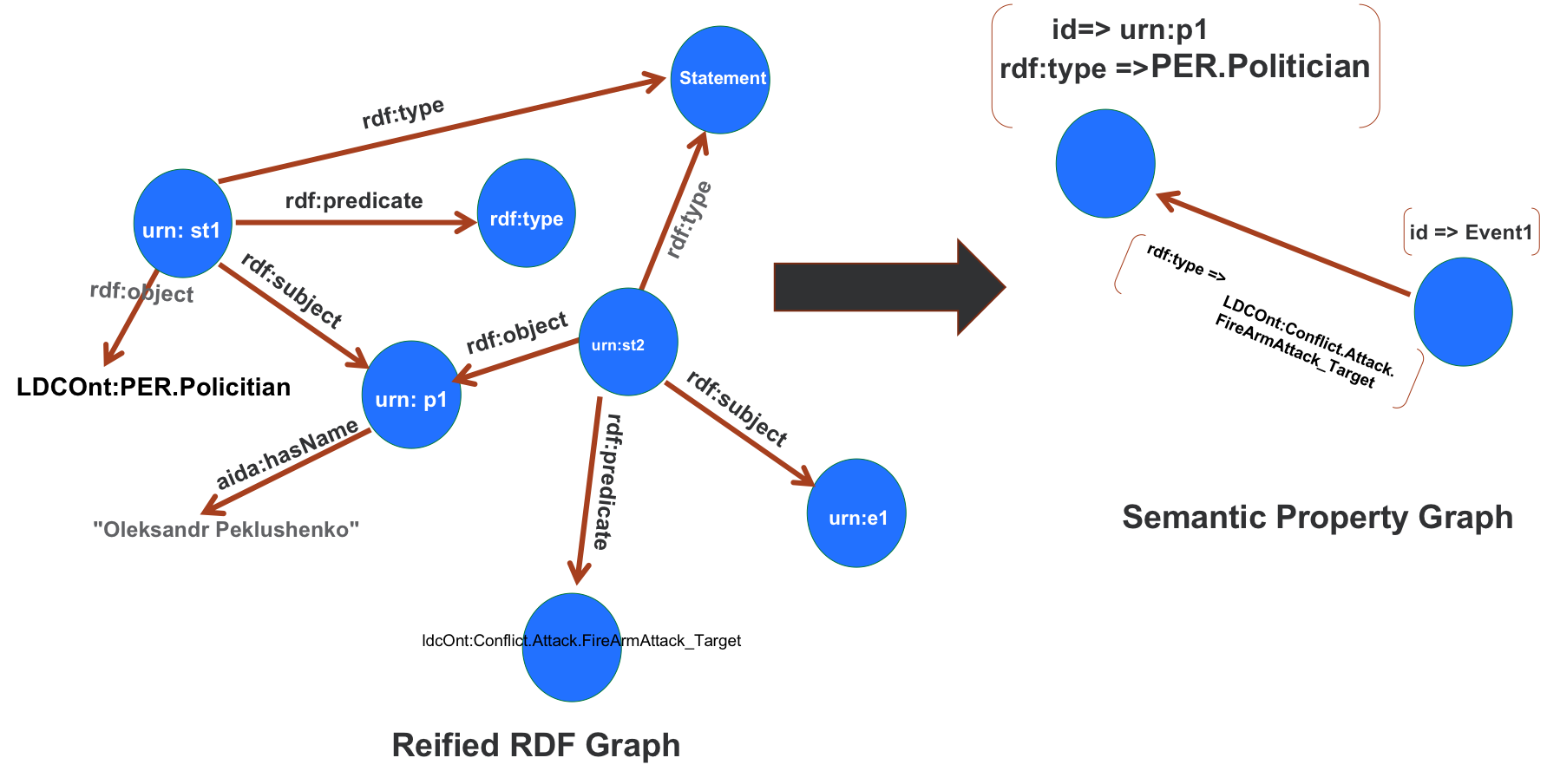}
	\caption{Semantic Property Graph Generation Example}
	\label{fig:spgexample}
\end{figure*}

SPG is a hybrid model to represent graph data. It is flexible and expressive, and at the same time provides efficient storage and faster traversal for graph analytic approaches. It is a property graph projection of reified RDF graphs and continues to use the underlying ontology used in the RDF graph model. Figure \ref{fig:spgexample} shows a nominal example of a reified RDF graph projection into a SPG. 

Most of the related work compares these two models in terms of graph database performance such as storage layout and query optimization \cite{das2014tale,angles2019directly}. Many extract, transform, load (ETL) processes have also been developed to convert RDF graphs into property graph format \cite{neo4jrdf,mitrdf}. The focus of such works is on specific RDF graphs and requires significant effort to apply those tools to graphs from other domains. We present a flexible framework to project RDF graphs into SPG. We continue to use RDF ontologies to define and validate node and edge properties in the projected graph. The framework supports different programming environments to select from while exporting RDF datasets. We also provide a cloud-based solution that uses Amazon Neptune to store and query large graphs. The framework provides a layer of abstraction to users and substantially reduces the effort required to export different RDF graphs into SPG.

\subsection{SPG Generation Framework}\label{sec:spggen}
\begin{figure}[H]
	\centering
	\includegraphics[width=.3\textwidth,height=.3\textheight]{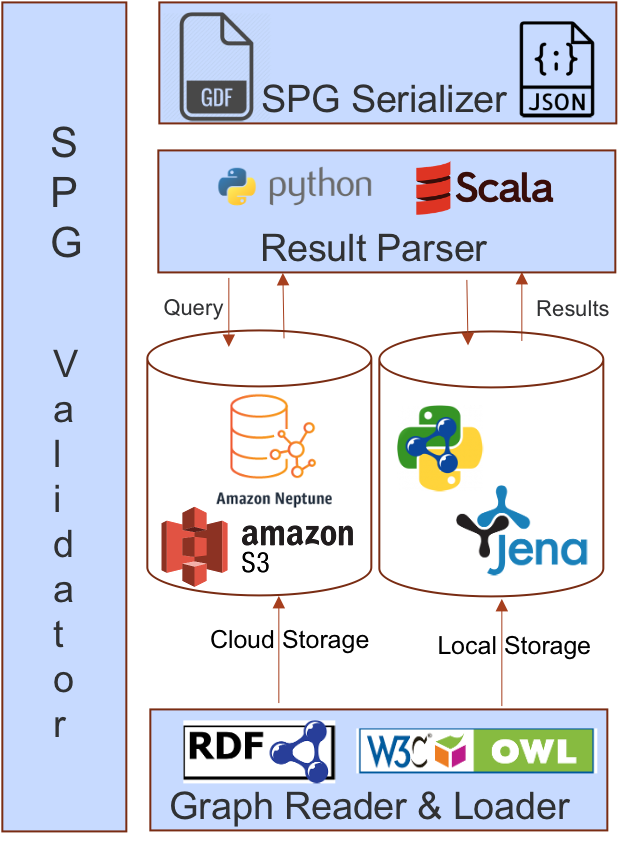}
	\caption{Semantic Property Graph Generation Framework}
	\label{fig:spgframwork}
\end{figure}

In this section, we present the framework to generate SPGs. We describe different components of the framework and the interfaces between them. Section \ref{sec:spgusecase} presents a Weapons of Mass Destruction (WMD) use case and the results. A major contribution of this work is a flexible cloud-scale framework that can be used by researchers and knowledge engineers in different domains to export the RDF graph into the SPG model. Figure \ref{fig:spgframwork} shows notional SPG generation architecture. The framework supports different programming environments such as Python and Scala. It defines the following components that interact with each other using programmable APIs. 
\begin{itemize}
	\item \textit{Graph Reader:} The RDF graph model can be serialized in different file formats which are logically equivalent. The choice of format is based on the requirements and capabilities of upstream graph generation components. Some of the frequently used RDF serialization formats include N-Triples, Turtle, and JSON-LD. The framework defines a graph reader component that supports reading various RDF serialization formats.
	\item \textit{Graph Loader:} The graph loader interacts with semantic storage and query engine and constructs a named graph. Two different modes of loader: \textit{local} and \textit{cloud} are developed to provide users the required flexibility based on the size of the input RDF graph. As the name suggest, \textit{local} mode uses an in-memory transient graph database based on Python RDFLib or Apache Jena libraries. Similarly, \textit{cloud} mode can access RDF graphs stored in Amazon Web Services (AWS) S3 buckets and can use Amazon Neptune as its persistent graph database. The Graph Loader provides an abstraction to the specific implementation of the underlying graph database.
	\item \textit{Semantic Storage and Query Engine:} As described above, the SPG generation framework supports two different modes of graph loading that correspond to two different types of graph storage and query engines. The Input RDF graph is loaded as a \textit{named graph}. All the \textit{local} named graphs are deleted at the end of SPG generation but we provide options to delete or keep the named graph loaded in the \textit{cloud} mode. This allows us to use generate varying projections of the RDF graph based on input queries.
	\item \textit{Result Parser:} The SPG generation framework provides an abstraction where the user needs to define only the projection query and its corresponding parser in order to generate the output semantic property graph. We define an interface with the required methods to generate valid SPG. We also provide reference implementations for example input RDF graphs and SPARQL queries. The Query parser interface supports tabular (i.e., CSV) or JSON result sets from the underlying query engine and needs to output a dictionary-based data structure defined by the framework. The output data structure identifies nodes and edges in the query result-set and is used by serializer and validator to generate a valid semantic property graph.
	\item \textit{SPG Serializer:} The SPG Serializer is a collection of methods to serialize dictionary based data structure generated by the Result Parser. We serialize SPG in a GDF file format which is built like a database table or a comma separated file (CSV). GDF supports attributes to both nodes and edges. A GDF file is divided into \textit{nodedef} and \textit{edgedef} sections. Each section starts with a header line that describes the names and types of each column in the section. Future work will add support for JSON-based SPG serialization.
	\item \textit{SPG Validator:} The SPG Validator works in combination with graph reader and result parser to validate the ontological type and data type of nodes and edges in the graph. 
\end{itemize}
\section{SPG Use Case}\label{sec:spgusecase} In this section, we present a WMD use case \cite{maa} that benefits from SPG. It develops mathematical and computational techniques for modeling adversarial activities observed in multisource datasets modeled as a semantic, temporal, and attributed graph. It identifies observable transactions to enable large-scale graph analytics tasks, including graph alignment and merging, subgraph detection, and subgraph matching. Input graphs are generated by the state-of-the-art NLP pipeline that identifies entities, events, and relationships in a corpus of text sources. The corpus is collected from a set of news articles, social media posts such as Reddit, transactional dataset such as Venmo, and domain-specific bibliographic information received from scientific publications. The input graphs use an ontology and are serialized as reified RDF graphs.

\begin{figure}
	\centering
	\includegraphics[width=.4\textwidth,height=.2\textheight]{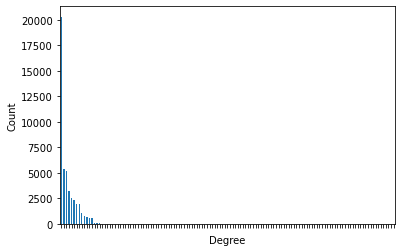}
	\caption{RDF Graph Degree Distribution}
	\label{fig:rdfdegdist}
\end{figure}

\begin{figure}
	\centering
	\includegraphics[width=.4\textwidth,height=.2\textheight]{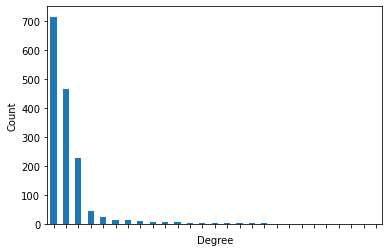}
	\caption{SPG Degree Distribution}
	\label{fig:spgdegdist}
\end{figure}

\begin{figure*}
	\centering
	\includegraphics[width=.8\textwidth,height=.3\textheight]{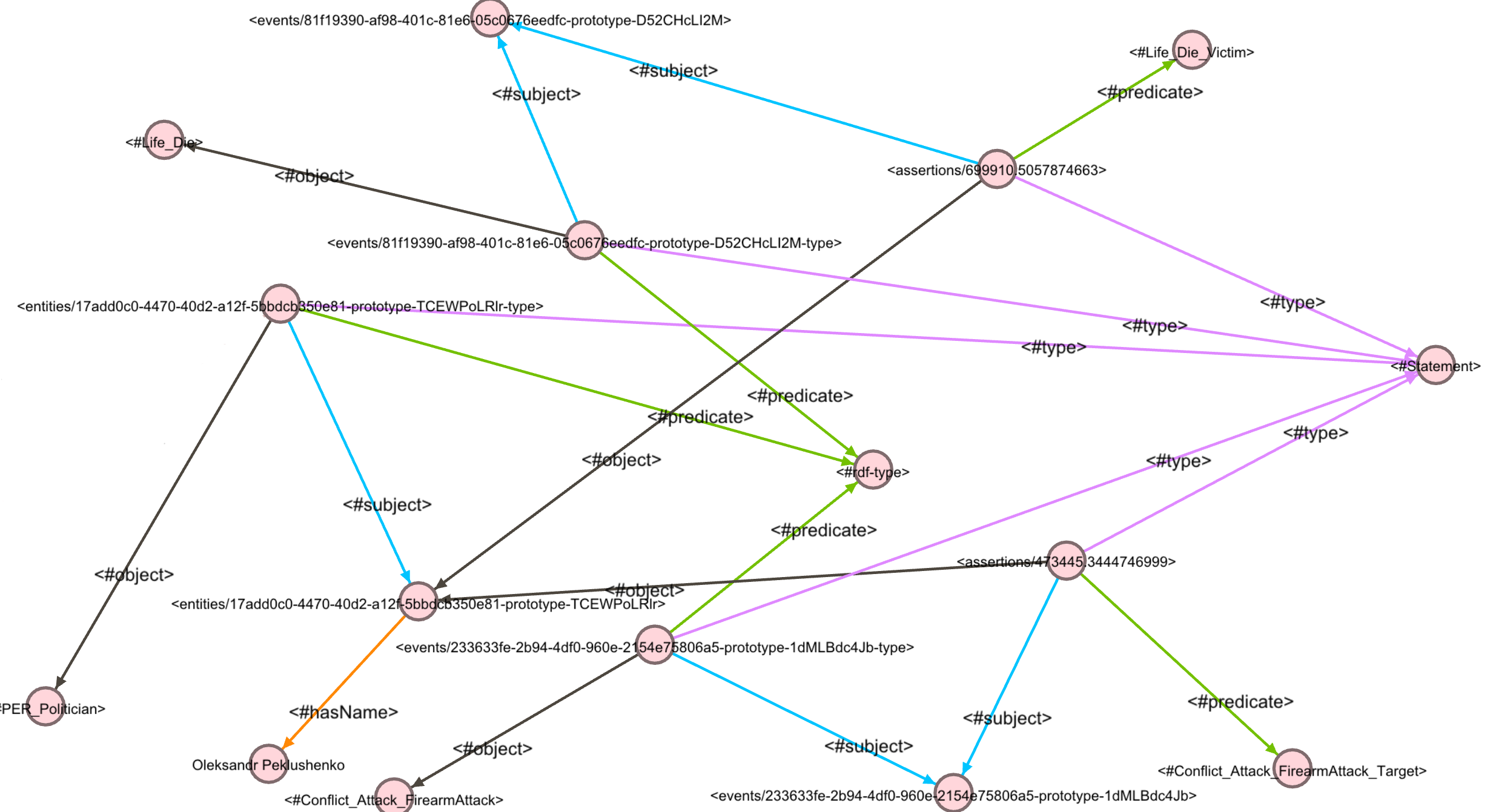}
	\caption{RDF Query Graph}
	\label{fig:rdfquerygraph}
\end{figure*}

\begin{figure}
	\centering
	\includegraphics[width=.3\textwidth,height=.2\textheight]{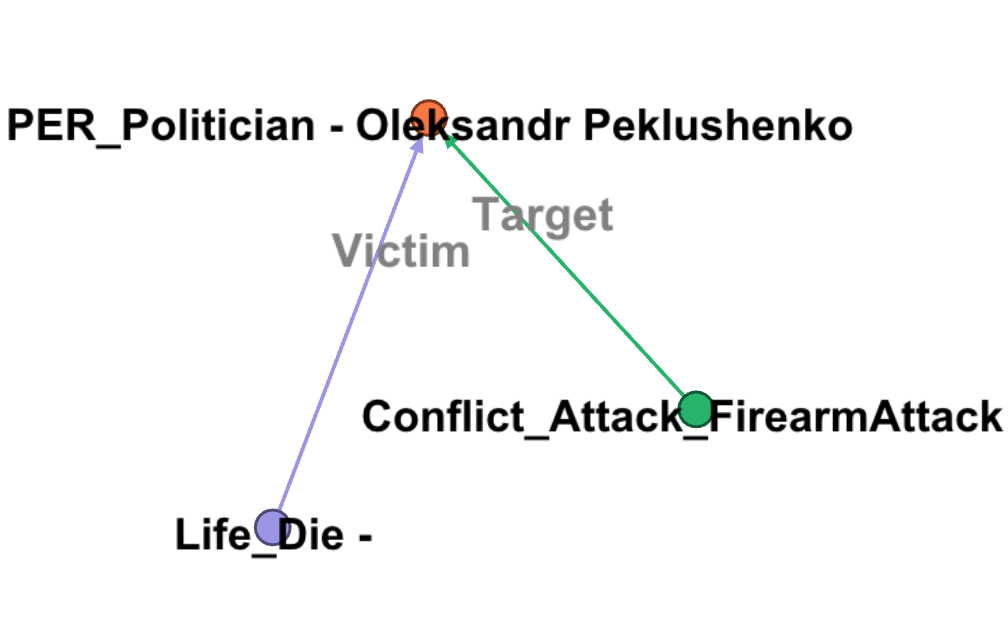}
	\caption{SPG Query Graph}
	\label{fig:spgquerygraph}
\end{figure}

Figure \ref{fig:rdfdegdist} and Figure \ref{fig:spgdegdist} show the degree distribution of theRDF and the SPG graphs generated using the NLP pipeline, respectively. RDF graphs show more skewed distribution in comparison with SPG. Since RDF defines a combined graph for data and metadata, all the high degree nodes correspond to ontological classes in the RDF graph such as \textit{Statement} and \textit{rdf:type}. Similarly, most of the edges define a \textit{Statement} in the RDF graph using \textit{rdf:subject},\textit{rdf:predicate}, \textit{rdf:object}. The capability to represent semantic information as part of the graph makes RDF an excellent choice for data exchange and applications that require inference capabilities. At the same time, this poses storage, traversal, and efficiency challenges for network analytic tasks such as network alignment, subgraph detection, and subgraph matching. Semantic Property Graph (SPG) provides compact, intuitive, and scalable graph representation for such tasks. Figure \ref{fig:spgquerygraph}  presents an attributed subgraph query expressed using SPG. This query is used as an input to the subgraph matching task and expresses a two-hop traversal using a wedge structure. In contrast, the equivalent RDF query is shown in Figure \ref{fig:rdfquerygraph} and is expressed as a larger subgraph of 20 edges and multiple instances of 3 hop subgraphs. Many RDF graph databases use an indexing technique to speed up the traversal, but the bloated nature of RDF query graphs make it difficult to use in graph analytic tasks. We extend the NLP pipeline by adding the SPG generation step and also generate a subgraph query library by generating query templates in GDF format.

\section{Conclusion}
We present the Semantic Property Graph (SPG), a hybrid approach to model attributed semantic graphs. SPG provides a flexible, configurable, and cloud-scale framework to project reified RDF graphs into the property graph model. SPG continues to use RDF ontologies but significantly reduces the graph structure and serialized file sizes without a loss of relevant information for graph analytic tasks. We have developed reference implementations in the Python and Scala programming environment and also support cloud-scale SPG generation using the Amazon Neptune graph database. Using a Weapons of Mass Destruction (WMD) use case, we present that researchers and knowledge engineers can benefit from the SPG framework.


\section*{Acknowledgment}
We thank the DARPA Modeling Adversarial Activity (MAA) program for funding this project under contracts HR0011728117, HR001178235, and HR0011729374.  The associated PNNL project number is 69986. A portion of the research was performed using PNNL Institutional Computing (PIC) at the Pacific Northwest National Laboratory.



%
\bibliographystyle{IEEEtran}
\bibliography{spg}

\end{document}